# Structural and Optical Properties of Spin-Coated $Mn_3O_4$ Thin Films of Different Coating Layers

**Vidit Pandey[1], Mohd Salman Siddiqui[2], Sandeep Munjal[3], Tufail Ahmad[1,*]**

[1]*Department of Physics, Aligarh Muslim University, Aligarh-202002, India.*
[2]*Department of Physics, Indian Institute of Technology Bombay, Mumbai-400076, India.*
[3]*National Forensic Sciences University, Goa-403401, India.*

**Abstract:**

In present study, Tetragonal $Mn_3O_4$ thin films of different coating layers of 2, 4, 6, 8, and 10 were prepared on the microscopic glass slides by the spin-coating method. X-ray diffraction, Raman spectroscopy, and ultra-violet visible spectroscopy were used to explore the structural and optical properties of $Mn_3O_4$ thin films. XRD patterns and Raman spectra confirm the phase of all films as tetragonal. The average crystallite size of films increased from ~ 17 to ~ 25 nm with increasing coating layers. Optical properties such as band gap, skin depth, cutoff wavelength, excitation coefficient, refractive index, etc, have been studied in detail by using ultra-violet visible spectroscopy. The direct optical band gap and refractive index of fabricated films were decreased as increasing the coating layers of $Mn_3O_4$. The refractive index for 2 and 4 coating layers were decreased as increasing the incident wavelength. While for 6, 8, and 10 coating layers, the values of refractive index remain constant with respect to incident wavelength. The optical analyses suggest that prepared $Mn_3O_4$ thin films can be used as a potential candidate for optical semiconductor devices including photovoltaic cells and optical sensors.

[*] Corresponding author. T. Ahmad
Tel.:+91-0571-2700920, ext.3610.
*E-mail address:* tufailahmadphys@gmail.com

## 1. Introduction

Over the past decades, the normal spinel structure ($AB_2O_4$, where A and B are metal ions [1]) transition metal oxides-based thin films have been explored because of its wide range of potential applications in the field of data storage [2], sensing [3] [4], catalysis [5], energy [6], antibacterial activities [7] [8], etc. Manganese oxide ($Mn_3O_4$) is one of the interesting spinel materials due to the most stable phase among other oxides of the Manganese atom, environmental friendly [9], and natural abundance [10]. In spinel tetragonal $Mn_3O_4$, the $Mn^{2+}$ and $Mn^{3+}$ ions occupy tetrahedral (A-site) and octahedral (B-site) positions, respectively. In previous reports, $Mn_3O_4$ thin film has been deposited on diverse substrates by numerous methods such as spray coating [11], spin coating [12] [13], chemical vapor deposition [14], pulsed laser deposition [15], atomic layer deposition [16], molecular beam epitaxy [17], etc. Among the various thin film deposition method, the spin coating method for the deposition of thin films is generally preferred over the other techniques as this method is cost-effective, and can be used to coat the film on a large substrate area with great control [18] [19].

The physical and chemical properties of tetragonal $Mn_3O_4$ thin films can be tuned by varying annealing temperatures [20], different molar concentrations of precursors during the reaction process [21], doping of other elements [22], reaction time [23], pressure [24], and incident photon energy [25]. Hence, it is used in various novel applications including resistive switching phenomenon-based memory devices (RRAM) [26] [27], cancer treatment [28], and water purification [29]. V. Pandey et al. [12] demonstrated bipolar resistive switching in $Al/Mn_3O_4/FTO$ RRAM devices and reported that spin-coated $Mn_3O_4$ thin films can be used for high-density data storage non-volatile memories. Y. Li et al. [30] investigated that $Mn_3O_4$ is a p-type semiconductor material with a direct optical band gap (~ 2-2.8 eV) and suggested that it can be utilized in photo-catalytic activities for water purification and water-splitting.

In metal oxide-based thin film, the optical engineering keep an importance to choose the best materials for optical parameters-based applications like water splitting [31], solar cells [32], photo-voltaic devices [33] [34], optical sensors [35], optical memories [2], fiber optics [36], optical switching [37] [38], etc. Therefore, these parameters are customized by altering other variables such as thickness of thin film, doping, taking concentration of metal's precursors, particle size, morphology, surface defects, crystalline phase, or number of coating layers [39]. In previously reported literature, the optical and structural parameters of $Mn_3O_4$ thin films on different substrate were tuned by numerous technique such as doping of other elements [40] [41] [42], varying concentration of precursors [25], different annealing temperature [43], varying thickness of $Mn_3O_4$ [44]. However, to best our knowledge, there is no published report on the tuning of optical parameters of $Mn_3O_4$ thin films on normal microscopic glass substrate by different spin-coating layers.

In this chapter, $Mn_3O_4$ thin films were prepared successfully by an eco-friendly spin coating method of different coating layer on normal transparent microscopic glass substrates. X-ray diffraction (XRD) and Raman spectroscopy were employed to investigate its structural's phase and optical phonon mode, respectively. Optical properties such as band gap, skin depth, cutoff wavelength, excitation coefficient, refractive index, etc, have been studied in detail by using ultra-violet visible (UV Vis.) spectroscopy. Optical analyses of fabricated $Mn_3O_4$ thin films suggest that it could be suitable for optical switching-based electronic devices, sensors, filters and photovoltaic cells.

## 2. Deposition and Experimental

### 2.1. Before Deposition

All the chemicals, precursors, materials, and substrates (Sigma-Aldrich) were of analytical grade and used without further purification in the process of synthesis. Doubly deionized (DI) water, which is the form of purified water having pH ~ 7, is used throughout the experiment. All glass vessels (volumetric beakers, measuring cylinders, flasks, test tubes, pipettes, etc.) and instruments were washed with acetone and DI water, and dried in a hot air oven at 90 $^{o}$C for half an hour, before using them for the synthesis process. A digital weighing machine, ultra-sonic bath, magnetic stirrer with a hot plate, spin-coating machine, digital hot air oven, and muffle furnace were employed in thin film preparation. Spin-Coating machine was used to deposit the Manganese's precursors on glass substrate.

### 2.2. Thin Film Deposition

$Mn_3O_4$ thin film on the microscopic glass substrate was deposited by a spin coating method. To fabricate the thin film, 0.05 M of manganese acetate ($Mn(CH_3CO_2)_2$•$4(H_2O)$) was dissolved in 2-methoxyethanol ($C_3H_8O_2$) and kept on a hot plate magnetic stirrer (450 rpm) at 45 $^{o}$C for 30 minutes (Step A). Few drops of liquid ammonia ($NH_3OH$) (0.2 ml) were added to maintain the pH of the solution ~ 10 (Step B). Meanwhile, the glass substrate was washed several times with ethanol and distilled water in an ultra-sonic bath. After that, the solution was spin coated on the 5 substrates at 3500 rpm for 15 seconds (Step C) followed by drying in an oven for 15 minutes at 125 $^{o}$C (Step D) andannealed the films in a muffle furnace at 460 $^{o}$C for 12 hours (Step E). Thus, 5 thin films were fabricated of different coating 2, 4, 6, 8, and 10 layers Figure 1 depicts the flow chart of the processes for fabrication of brown color films on the glass substrate.

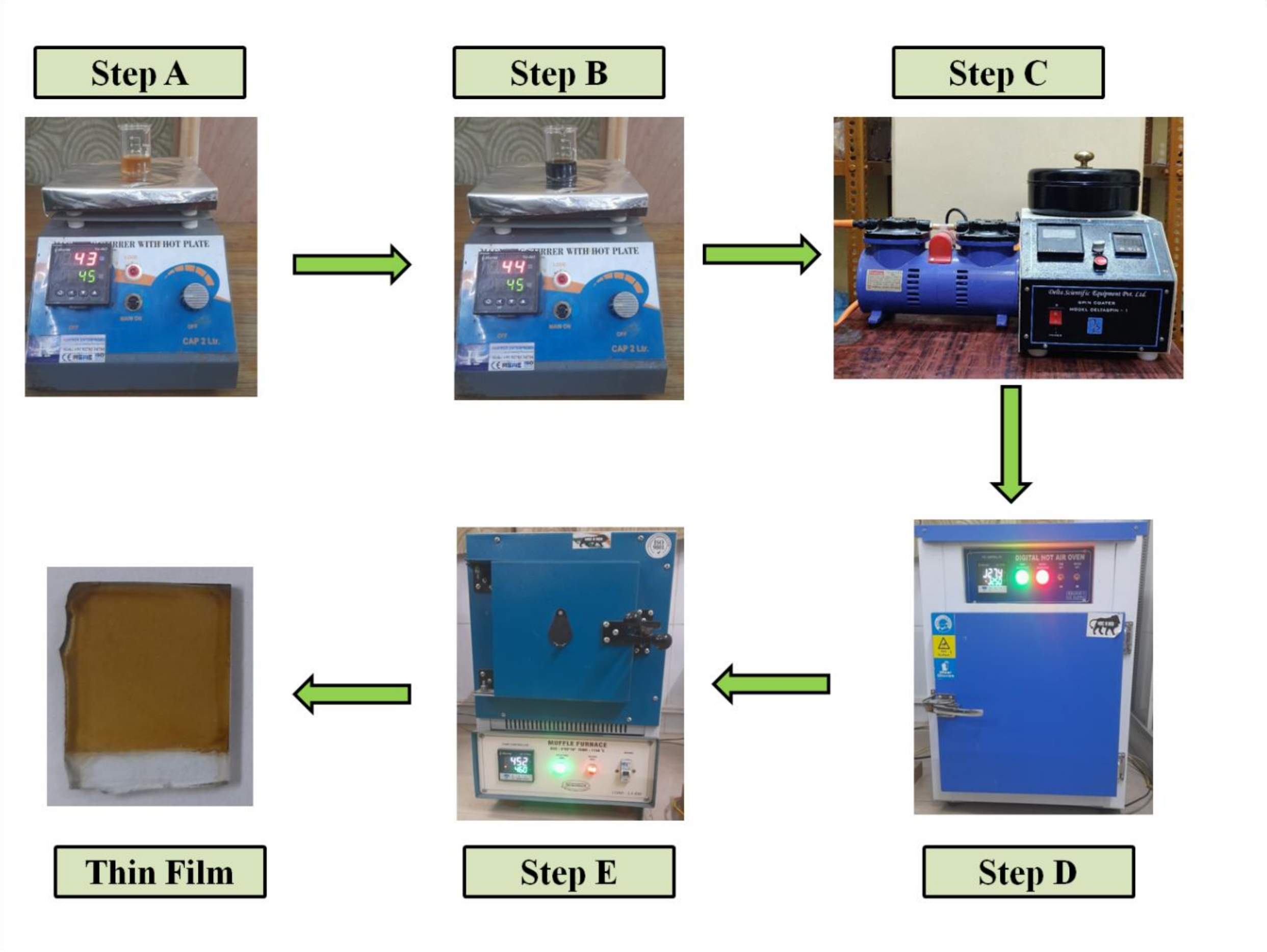

**Fig. 1.** The flow chart of the process of deposition of $Mn_3O_4$ thin film on the microscopic glass substrate. Steps A, B, C, D, and E depict the formation of Manganese precursor solution, adding few drops of ammonia, spin coating process of solution, drying thin film in an oven, and annealing film in furnace, respectively. The finally obtained a light brown color thin films' image.

## 2.3. Experimental

The synthesized $Mn_3O_4$ thin films were characterized by XRD and Raman. X-ray diffraction (XRD) measurements were performed at room temperature using Shimadzu LabX XRD-6100 X-ray diffractometer equipped with X-rays (Cu-Kα) of wavelength ~ 1.542 Å. Raman spectrographs of the films were recorded with the help of Raman spectroscopy (Renishaw Centrus, model:  1C4A91, laser: 532 nm edge with power 5%, and grating: 2400I/nm) at room temperature. The optical properties of the sample have been studied using Lambda UV vis. DRS (Perkin Elmer, USA) in the range from 200 to 800 nm at room temperature.

## 3. Results and Discussion

### 3.1. X-Ray Diffraction (XRD)

#### 3.1.1. Phase Analyses

XRD analyses have been carried out to characterize the phase of the fabricated thin films, which were recorded in the range of diffraction angle (2θ) from $15^o$ to $70^o$ at room temperature (Fig. 2). XRD patterns of all thin films belong to tetragonal symmetry of $Mn_3O_4$ of space group I41/amd(#141) [16] and match with the joint committee on powder diffraction standards (JCPDS) card No.- 01-089-4837 [45]. No other peaks have been observed, which confirm the presence of pure phase of tetragonal $Mn_3O_4$. The first peak (101) was indexed at ~ $18.38^o$, followed by (112), (103), (004), (303), (215), and (411) at ~ $29.01^o$, ~ $32.99^o$, ~ $37.34^o$, ~ $55.28^o$, ~ $60.63^o$, and ~ $65.33^o$, respectively in all thin films of tetragonal $Mn_3O_4$ (Fig. 6.3(f)) [46].

In XRD patterns, a baseline hoop was also observed in the range of diffraction angle from ~ $21^o$ to ~ $26^o$ in all films, which refers the amorphous phase of microscopic glass slide [47]. A. Kocyigit [48] also reported the similar XRD baseline hoop in XRD pattern of bare $Mn_3O_4$ on glass slide. The XRD pattern of metal oxide based-thin film on glass substrate is noisy due to the dominancy of amorphous phase of glass in comparison to same metal oxide based-thin film on other substrate such as FTO, STO, Silicon, Aluminum, etc. [43]. On the other hand, the XRD pattern also noisy at low molar concentration of Manganese precursors (~ 0.05 M) [25]. It is noticed that in 2 coating layer of $Mn_3O_4$ thin film (Fig. 2(a)), there is no appearance of the XRD peaks of tetragonal $Mn_3O_4$. As increasing the spin-coating layers (from 4 coatings in this study, as shown in Fig. 2 (b)), the crystalinity of $Mn_3O_4$ increased. In 8 and 10 coating layers as shown in Fig. 2(d) and Fig. 2(f), respectively, the sharp XRD peaks were indexed. The intensity of amorphous hoop of the microscopic glass slides slightly decreased as increasing the coating layers (Fig. 2(f)).

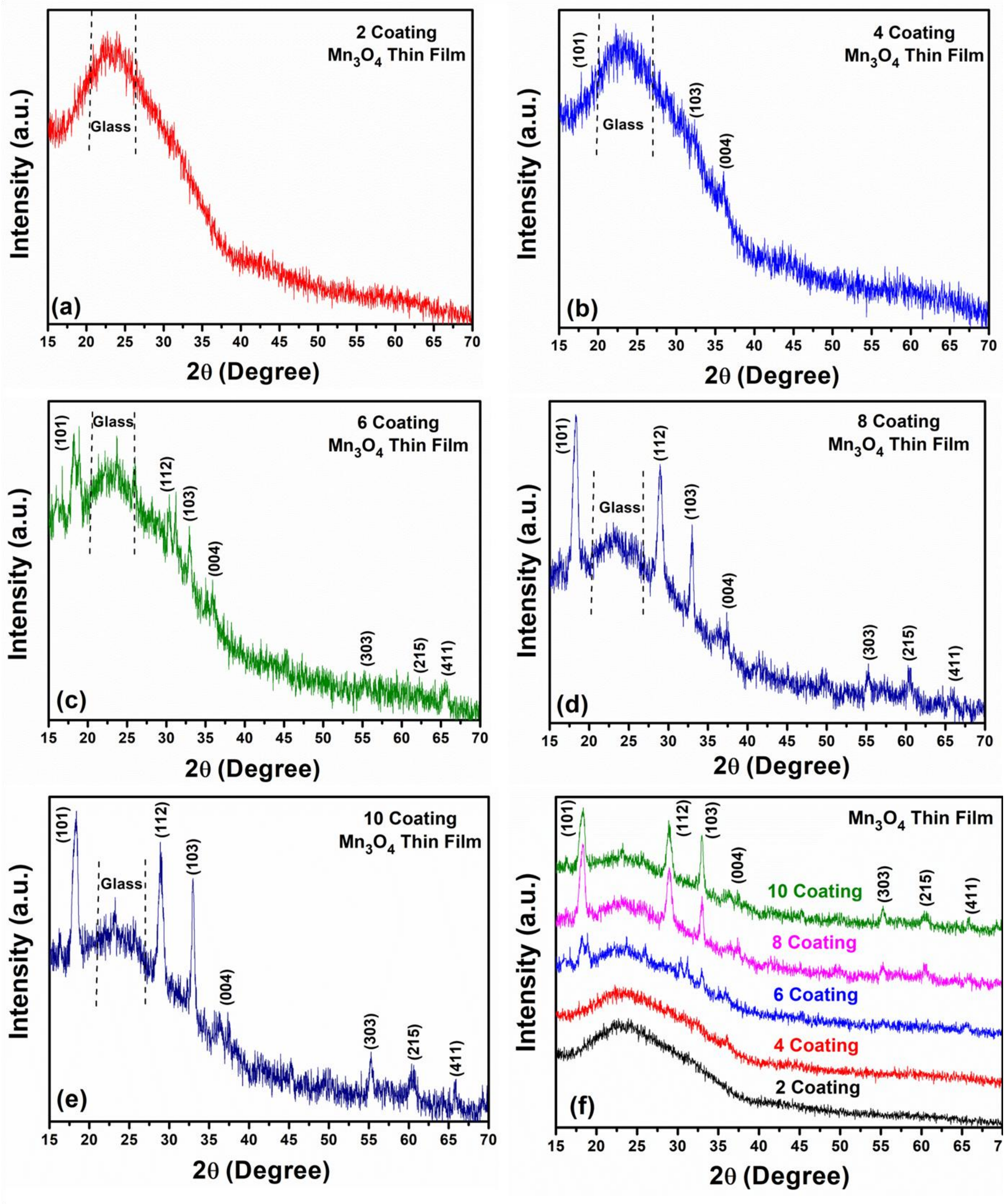

**Fig. 2.** XRD patterns of the tetragonal $Mn_3O_4$ thin films on microscopic glass slides of (a) 2 coating, (b) 4 coating, (c) 6 coating, (d) 8 coating, and (e) 10 coating, respectively at room temperature. (f) all XRD patterns of $Mn_3O_4$ thin films.

#### 3.1.2. Crystallite Size and Dislocation Density

In order to calculate the average crystallite size, the Debye-Scherrer mathematical formula was employed as shown in Eq. 1 [49]

$$t = \frac{k\lambda}{\beta \cos\theta} \quad (1)$$

Here, t is the average crystallite size, k is the Scherrer's constant (~ 0.89) [50], λ (1.5402 Å) is the wavelength of X-rays used in XRD, β is full width at half maximum (FWHM), and θ represent the Bragg's angle. The calculated values of crystallite size of 4, 6, 8, and 10 spin-coating layers $Mn_3O_4$ thin films are tabulated in Table 1. As increasing the number of coating of $Mn_3O_4$, the value of crystallite size has been enhanced from ~ 17.23 to ~ 23.30 nm (Fig. 3(a)).

Dislocation density gives the idea to know the irregularity in the crystal structure of nanoparticles. The mathematical expression for dislocation density can be expressed as [51]

$$\delta = \frac{1}{t^2} \quad (2)$$

here, δ is dislocation density in lines/$m^2$. t is the average crystallite size in nm (obtained from Eq. 1). The average value of dislocation density of fabricated $Mn_3O_4$ thin films (see Table 1) was decreased as increasing the number of coatings as shown in Fig. 3(b). The less values of dislocation density of the prepared thin film show less irregularity.

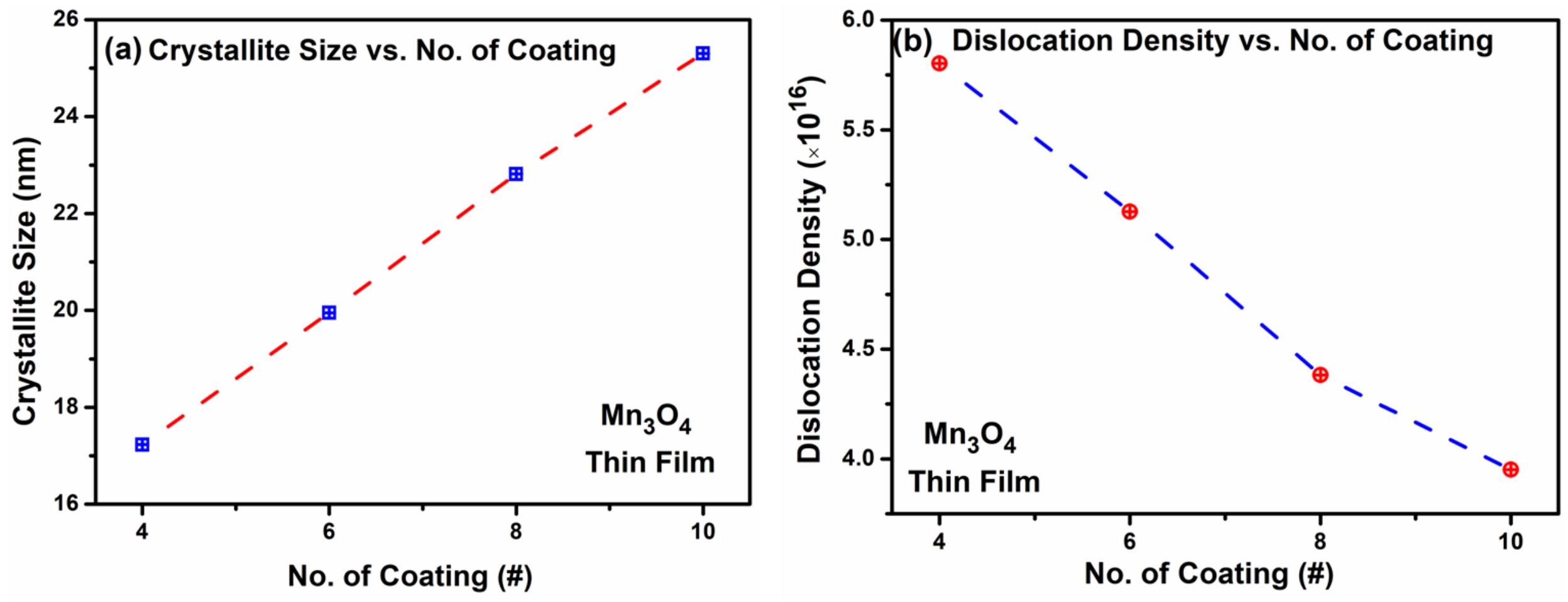

**Fig. 3.** The comparison of (a) crystallite size and (b) dislocation density with spin-coating layers of $Mn_3O_4$ thin films.

| Sr. No. | No. of Coating | Phase | Crystallite Size (nm) | Dislocation Density (lines/m$^2$) |
|---|---|---|---|---|
| 1- | 10 | Tetragonal | ~ 25.3045 | ~ $3.951\times10^{16}$ |
| 2- | 8 | Tetragonal | ~ 22.8163 | ~ $4.382\times10^{16}$ |
| 3- | 6 | Tetragonal | ~ 19.9502 | ~ $5.127\times10^{16}$ |
| 4- | 4 | Tetragonal | ~ 17.2304 | ~ $5.803\times10^{16}$ |

**Table 1.** Structural parameters of $Mn_3O_4$ thin films of different coating layers.

### 3.2. Raman Analyses

To further confirm the phase of $Mn_3O_4$ thin films, Raman spectroscopy was performed in the range of Raman wavelength from 150 to 1200 cm$^{-1}$ at room temperature using a laser of excitation laser wavelength 512 nm as shown in Fig. 4. Raman spectroscopy is the surface-sensitive technique to explore the optical phonon modes, crystal structure, phase transition, vibrational characteristics of spinel structure [52]. Among the various optical phonon modes, four Raman active modes $E_g$, $T_{2g}(1)$, $T_{2g}(2)$, and $A_{1g}$ for tetragonal $Mn_3O_4$ were obtained, which agree with previously reported studies [53] [54]. In obtained Raman spectrograph, $A_{1g}$ is a strong peak and is attributed to the stretching vibration of $Mn^{2+}$-O at the tetrahedral site of the spinel structure [55]. The peak $T_{2g}(1)$ refers to the stretching vibration of $Mn^{3+}$-O at the octahedral site [48], while the $T_{2g}(2)$ peak is related to the presence of $Mn^{4+}$ ions at the octahedral site [21]. It is pointed out that XRD patterns do not give any idea about the presence of $Mn^{4+}$ ions in the deposited $Mn_3O_4$ thin films. Raman spectrums' results of the prepared thin films are consistent with XRD analyses.

Apart from the optical phonon modes of the tetragonal $Mn_3O_4$, two Raman peaks at ~ 577 cm$^{-1}$ and ~ 1100 cm$^{-1}$ are observed in Raman spectrographs, which belongs to glass substrate [47] [56]. It is noticed that the Raman intensity of Raman peaks for 2 coating layer of $Mn_3O_4$ (Fig. 4 (a)) is higher than that of 10 coating film (Fig. 4(b)). As increasing the coating layers, the dominancy of glass Raman peaks decreases due to enhancement of the concentration of $Mn_3O_4$ layers. In 10 coating layers-based $Mn_3O_4$ thin film, the Raman spectrograph is similar to the $Mn_3O_4$ nanoparticles as shown in previous studies [46] [57].

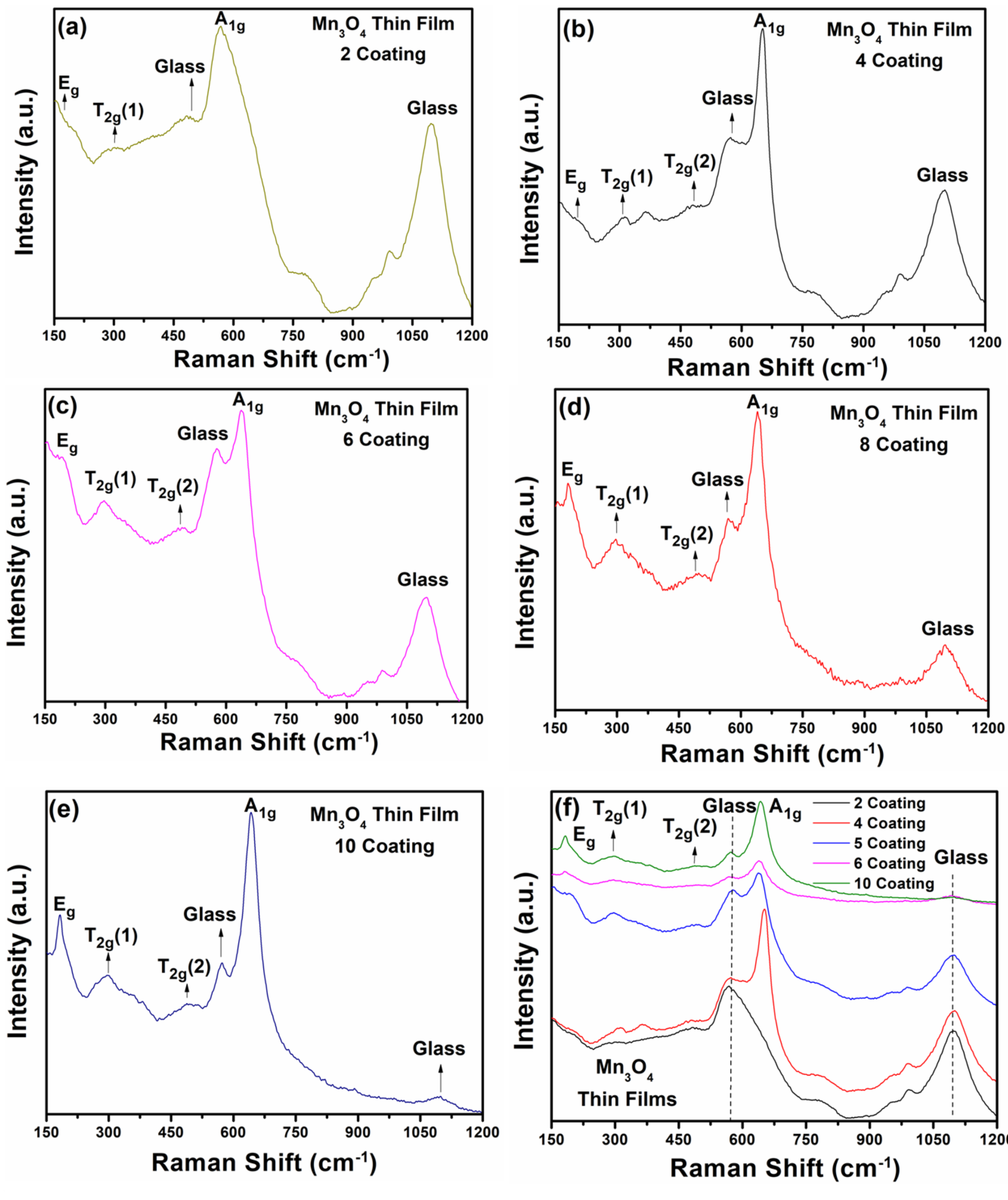

**Fig. 4.** The Raman spectrographs of the tetragonal $Mn_3O_4$ thin films on microscopic glass slides of (a) 2 coating, (b) 4 coating, (c) 6 coating, (d) 8 coating, and (e) 10 coating, respectively at room temperature. (f) All Raman spectra of $Mn_3O_4$ thin films in a single graph

### 3.3. Optical Analyses

In order to optical investigation of fabricated $Mn_3O_4$ thin films, UV Vis. (Perkin Elmer) was employed in the range of wavelength from 200 to 850 nm at room temperature. It is one of the important tools to determine the optical band gap, skin depth, excitation coefficient, refractive index, optical cut-off frequency, etc. In the present chapter, we focus on the detailed optical properties of spin-coated tetragonal $Mn_3O_4$ thin films for potential applications such as optical modulation [58], optical switching [2], optical memory devices [59], etc. . Therefore, the present studies target to explore numerous optical parameters of $Mn_3O_4$ thin films from reflectance, transmittance, and absorbance measurements.

#### 3.3.1. Optical Band Gap

The optical band gap (also known as the energy band gap) is the difference between the valance band and conduction band of a solid material. In other words, it is the state of energy in the solid, where no electronic states exit. Therefore, the optical band gap is an important term to divide the materials among conductors, semiconductors, and insulators. The Tauc's relation [60] and Kubelka-Munk (KM) model [31] were employed to estimate the optical band gap of $Mn_3O_4$ thin films. In the KM model, the KM function (F(R)) has a direct relation with reflectance (R) in the following way (Eq. 3)

$$F(R) = \frac{1 - R^2}{2R} \quad (3)$$

The calculated value of F(R) was applied in Tauc's relation (Eq. 4)

$$F(R) \cdot h\nu \ = \ A[h\nu - \ E_g]^n \quad (4)$$

where, hν and A are the energy of UV radiation and a constant, respectively. The value of n depends on type of the transitions. For indirect forbidden, direct forbidden, indirect allowed, and direct allowed transitions, the value of n is 3/2, 3, 1/2, and 2, respectively [61]. The Tauc's plots of all thin films to estimate the average value of direct optical band gap are shown in Fig. 5. As increasing the spin coating layers of the $Mn_3O_4$, the direct optical band gap decreased from ~ 2.56 to 1.85 eV due to increasing the thickness [44] or concentrations of $Mn_3O_4$ [25] (Fig. 5(f)). Increasing crystallite size of metal oxides may be one of the reasons of decreasing the value of direct optical band gap [62]. In our case, crystallite size increased as increasing the spin-coating layers of $Mn_3O_4$ (Fig. 3(a)). The obtained values of direct optical band gaps of the films belong to the visible region [63], which confirms that tetragonal $Mn_3O_4$ may have potential to employee in solar cell or water splitting applications [32].

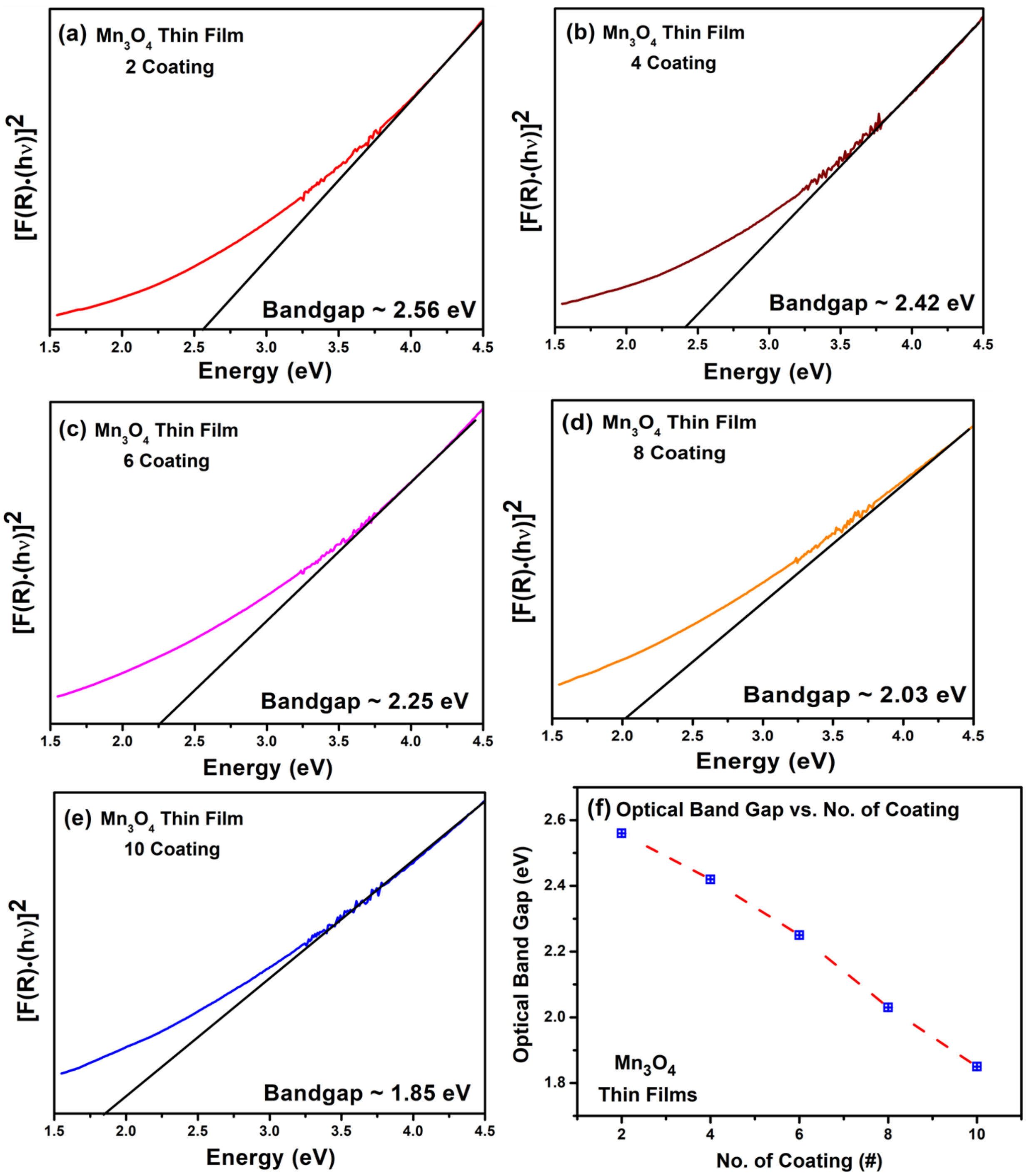

**Fig. 5.** Tauc's plot of the $Mn_3O_4$ thin films (a) 2 coating, (b) 4 coating, (c) 6 coating, (d) 8 coating, and (e) 10 coating, respectively at room temperature. (f) Direct optical band gap vs. no. of spin coating layers of tetragonal $Mn_3O_4$ thin films.

### 3.3.2. Skin Depth and Cutoff Wavelength

The optical skin depth (δ) is defined as the decay of electromagnetic waves when it penetrates the film surface or solid. It is also known as penetration depth and is inversely proportional to the absorption coefficient (α) [64]. It suggests how deep the current, electric, and magnetic fields can penetrate the conducting surface of a solid. Figure 6(a) shows that the optical skin depth of 2 and 4 spin coating-based $Mn_3O_4$ thin film decreases with increasing the incident photon energy. While, this value for 6 and 8 spin coating-based $Mn_3O_4$ thin films remains constant upto 5.5 eV and above this skin depth value slightly increased. In case of 10 spin coating-based film, the optical skin depth decreases upto 5.2 eV and after this abruptly increased due to the high concentration of tetragonal $Mn_3O_4$ layers.

The wavelength at which the value of skin depth becomes zero is known as cutoff wavelength ($\lambda_{Cutoff}$) or threshold wavelength. In order to calculate the $\lambda_{Cutoff}$ of incident radiation, the UV Vis. spectra of all thin films were analyzed with the help of the following Eq. 5 [64].

$$\left(\frac{A}{\lambda}\right)^2 = G\left(\frac{1}{\lambda} - \frac{1}{\lambda_{cutoff}}\right) \tag{5}$$

here, A, G, and λ are absorbance, optical empirical constant, and wavelength of incident photon, respectively. The obtained values of G of all films are tabulated in Table 2. The average value of $\lambda_{Cutoff}$ of all films was estimated by the plot between $(A/\lambda)^2$ and $(1/\lambda)$ as shown in Fig. 6.6(b), which is found ~ 655 nm (see Table 2 for $E_{Cutoff}$ for all fabricated films). R. Vignesh et al. [25] also observed the similar values of $\lambda_{Cutoff}$ for spray pyrolysis deposition-based tetragonal $Mn_3O_4$ thin film on microscopic glass substrate.

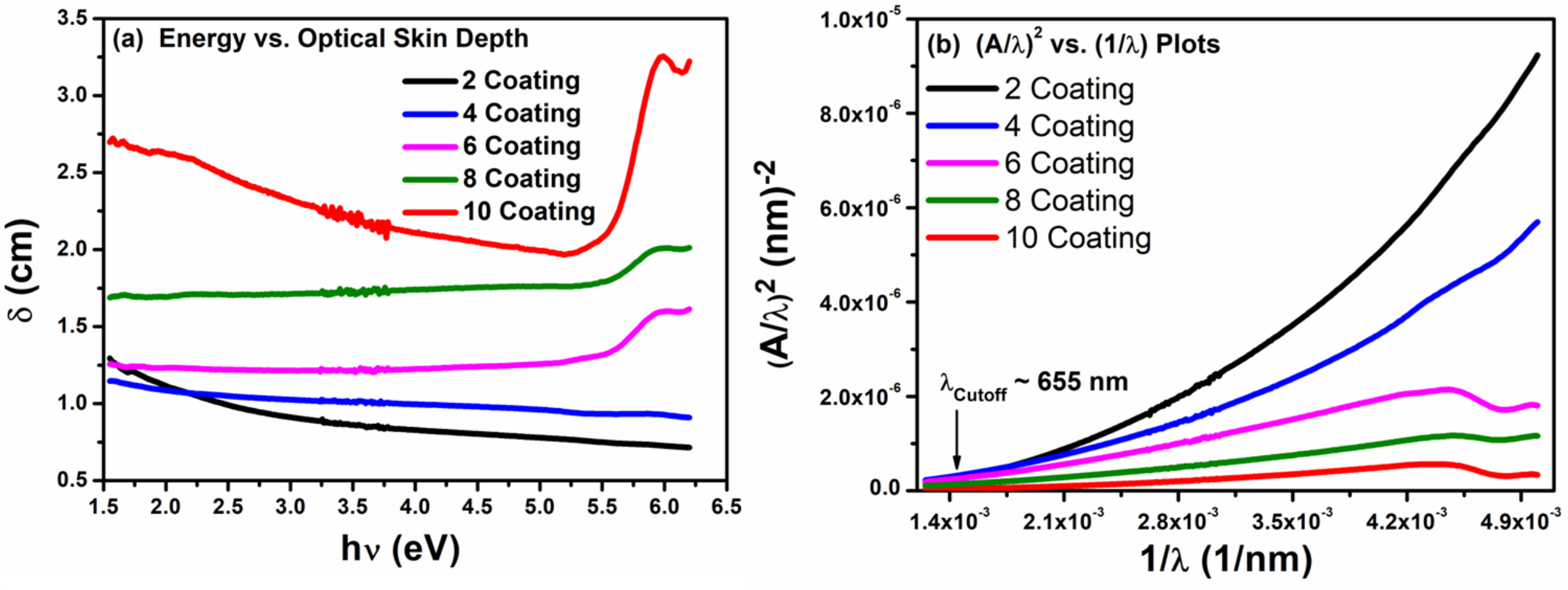

**Fig. 6.** (a) The energy vs. optical skin depth analysis of fabricated $Mn_3O_4$ thin films of different coating layers at room temperature. (b) The plot between $(A/\lambda)^2$ and $(1/\lambda)$ to estimate the cutoff wavelength and optical empirical constant for all fabricated thin films.

### 3.3.3. Refractive Index and Excitation Coefficient

In the optical engineering of thin films, both the extinction coefficient ($k_{ec}$) or attenuation constant and refractive index ($n_{ri}$) are important terminologies, and they vary with respect to incident wavelength or frequency [25]. The determination of the values of $n_{ri}$ and $k_{ec}$ is important for the fabrication of optical devices such as modulation, filters, switches, etc. When electromagnetic waves travel through a material's medium, the loss of dielectric energy has occurred in scattering, generation of phonons, or formation of free carrier absorption. Therefore, both $n_{ri}$ (as the real part) and $k_{ec}$ (as the imaginary part) of the materials exist in complex mathematical functions as given in Eq. 6.

$$n^* = n_{ri} - ik_{ec} \quad (6)$$

Where, n* represents the complex form of $n_{ri}$. The $n_{ri}$ and $k_{ec}$ can be analyzed by using reflectance and absorption spectra with the help of the following equations (Eq. 7 and 8) [65]

$$n_{ri} = \left[\frac{1+\sqrt{R}}{1-\sqrt{R}}\right] \quad (7)$$

$$k_{ec} = \frac{\alpha\lambda}{4\pi} \quad (8)$$

here, R is the reflectance of the fabricated thin films. Figure 7(a) depicts the refractive index of the 2, 4, 6, 8, and 10 spin coated layers of tetragonal $Mn_3O_4$ with varying incident wavelength at room temperature. The value of $n_{ri}$ of the 2 spin coating layers-based of $Mn_3O_4$ thin film is ~ 1.77 in visible regions, whereas it decreases upto ~ 1.42 in ultra-violet regions. The 4 spin coated layers-based $Mn_3O_4$ thin film also shows slightly variation in the value of $n_{ri}$ from ~ 1.58 to ~ 1.52 as increasing the incident wavelength. On the other hand, the refractive index of 6, 8, and 10 spin coated layers-based $Mn_3O_4$ thin films are stable in both visible and ultra-violet region. Stability of refractive index of metal oxide-based thin films with respect to incident wavelength or frequency is one of the challenging tasks for optics-based industries and our fabricated thin films (6, 8, and, 10 spin coated layers) show excellent stability, which is suitable to manufacture optical filters or frequency-cutter [66] [67] [68] [69] [70]. It is noticed that the refractive index of $Mn_3O_4$ thin films decreases from ~ 1.77 to ~ 1.19 as increasing the spin-coated layers and refractive index show dependency on number of spin-coated layers rather than incident wavelength or energy. The values of $n_{ri}$ of all fabricated thin films are tabulated in Table 2.

On the other side, the value of $k_{ec}$ of all fabricated thin films increases as increasing the incident wavelength from 250 to 800 nm at room temperature as shown in Fig. 7(b). In visible region, the values of $k_{ec}$ of all films are in the range of ~ 0.5-1×$10^{-5}$. While these values approach to ~ 1.5-2.25×$10^{-5}$ in ultra-violet region.

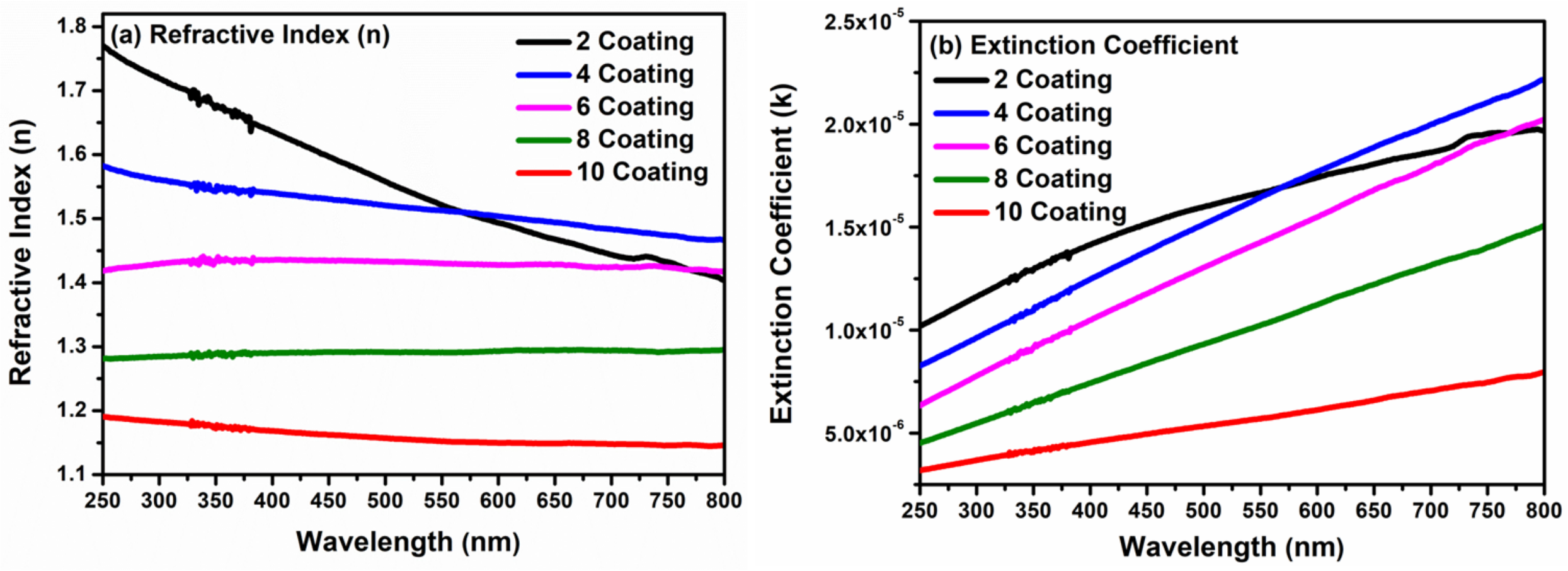

**Fig. 7.** The refractive index(a) and excitation coefficient (b) with respect to the incident wavelength of fabricated 2, 4, 6, 8, and, 10 spin coated layers-based $Mn_3O_4$ thin films.

| Sr. No. | Optical Parameters | 2 Coating | 4 Coating | 6 Coating | 8 Coating | 10 Coating |
|---|---|---|---|---|---|---|
| 1- | Band Gap (eV) | ~ 2.56 | ~ 2.42 | ~ 2.25 | ~ 2.03 | ~ 1.85 |
| 2- | Empirical constant | ~ $1.95\times10^{-3}$ | ~ $1.21\times10^{-3}$ | ~ $5.79\times10^{-4}$ | ~ $3.11\times10^{-4}$ | ~ $1.43\times10^{-4}$ |
| 3- | $\lambda_{Cutoff}$ (nm) | ~ 674 | ~ 663 | ~ 655 | ~ 650 | ~ 644 |
| 4- | $E_{cutoff}$ (eV) | ~ 1.84 | ~ 1.87 | ~ 1.89 | ~ 1.91 | ~ 1.92 |
| 5- | Refractive Index ($n_{ri}$) | ~ 1.77 - ~ 1.42 | ~ 1.58 - ~ 1.52 | ~ 1.42 | ~ 1.28 | ~ 1.91 |

**Table 2.** Optical parameters of $Mn_3O_4$ thin films of different coating layers.

The obtained optical parameters of the spin-coated $Mn_3O_4$ thin films are compared with the existing literature and tabulated in Table 3.

| Sr. No. | Deposition Method | Precursors | Substrate | Optical Band Gap (eV) | Tuned by | Effect on Band Gap | References |
|---|---|---|---|---|---|---|---|
| 1- | SILAR Technique | 0.1 M Manganese Nitrate | Soda Lime Glass | ~ 2.06-1.71 | Doping of Copper (Cu) | Decrease with Increasing Concentration of Cu | [40] |
| 2- | Chemical Spray Pyrolysis | 1.97 g Manganese Chloride in | Glass | ~ 2.89-2.87 | Doping of Barium (Ba) | Remains Same | [41] |
| | | 100 ml DI water | | | | | |
| 3- | Nebulized Spray Pyrolysis | 0.04-0.2 M of Manganese Chloride | Microscopic Glass Slides | ~ 2.93-2.43 | Varying Concentration of Precursor | Decrease with Increasing Concentration | [25] |
| 4- | Spin Coating | 5.0 m mole of Manganese Acetate | Glass | ~ 1.34-1.02 | Different Annealing Temp. (350 – 550 $^{o}$C) | Decrease with Increasing Temp. | [43] |
| 5- | Spin Coating | 5.0 m mole of Manganese Acetate | Silicon | ~ 3.04-2.74 | Different Annealing Temp. (350 – 550 $^{o}$C) | Decrease with Increasing Temp. | [43] |
| 6- | SILAR Technique | 0.15 M of Manganese Nitrate | Soda Lime Glass | ~ 2.20-2.38 | Doping of Lead (Pb) | Increase with Increasing concentration of Pb | [42] |
| 7- | Chemical Spray Pyrolysis | Manganese Accetate | Glass | ~ 3.83-2.85 | Different Thickness (200-350 nm) | Decrease with Increasing Thickness | [44] |
| 8- | Spray Pyrolysis | 0.1 M Manganese Nitrate | Si/Glass | ~ 2.33-2.13 | Doping of Gold (Au) | Decrease with Increasing Concentration of Au | [48] |
| 9- | SILAR Technique | 0.1 M Manganese Nitrate | Soda Lime Glass | ~ 2.05-1.73 | Doping of Zinc (Zn) | Decrease with Increasing Concentration of Zn | [71] |
| 10- | Spin Coating | 0.05 M of Manganese acetate | Microscopic Glass Slides | ~ 2.56-1.85 | Spin Coating Layers (2 to 10 Layers) | Decrease with Increasing Spin-Coated Layers | This Work |

**Table 3.** Comparison of the tuning of optical parameters of spin-coated $Mn_3O_4$ thin film with previously reported studies.

## 4. Conclusion

In present work, $Mn_3O_4$ thin films of different spin coating layers have been prepared by eco-friendly and economically spin-coating method, successfully. XRD patterns of all prepared films confirm the presence single phase of tetragonal $Mn_3O_4$. The average crystallite size of films increased from ~ 17 to ~ 25 nm with increasing coating layers. Raman spectra of all films authenticated the XRD analyses and the sharp peak $A_{1g}$ (~ 657 $cm^{-1}$) was dedicated to $Mn^{2+}$-O at tetrahedral sites. The intensity of Raman peaks of microscopic glass substrate was high in 2 coating layers-based thin film and it decreased or vanished in 10 coating layers-based thin film. The value of direct optical band gaps were found in range from ~ 2.56 to ~ 1.85 eV and it decreased with increasing the coating layers. The average value of cutoff frequency of all films was obtained as ~ 655 nm. Refractive index of the $Mn_3O_4$ thin films were shown variation for 2 and 4 spin-coated layers and stable for 6, 8, and 10 spin-coated layers. These optical analyses suggest that fabricated $Mn_3O_4$ thin film on a glass substrate can be treated as a potential candidate for optical semiconductor devices including photovoltaic cells, optical sensors and filters.